\documentclass[12pt]{article}
\usepackage{amsmath,amsfonts,amssymb}

\begin{document}
\thispagestyle{empty}

\def\theequation{\arabic{section}.\arabic{equation}}
\def\a{\alpha}
\def\b{\beta}
\def\g{\gamma}
\def\d{\delta}
\def\dd{\rm d}
\def\e{\epsilon}
\def\ve{\varepsilon}
\def\z{\zeta}
\def\B{\mbox{\bf B}}\def\cp{\mathbb {CP}^3}

\newcommand{\h}{\hspace{0.5cm}}

\begin{titlepage}
\vspace*{1.cm}
\renewcommand{\thefootnote}{\fnsymbol{footnote}}
\begin{center}
{\Large \bf Giant magnon-like solution in $Sch_5\times S^5$}
\end{center}
\vskip 1.2cm \centerline{\bf Changrim  Ahn and Plamen Bozhilov
\footnote{On leave from Institute for Nuclear Research and Nuclear
Energy, Bulgarian Academy of Sciences, Bulgaria.}}

\vskip 10mm

\centerline{\sl Department of Physics} \centerline{\sl Ewha Womans
University} \centerline{\sl DaeHyun 11-1, Seoul 120-750, S. Korea}
\vspace*{0.6cm} \centerline{\tt ahn@ewha.ac.kr,
bozhilov@inrne.bas.bg}

\vskip 20mm

\baselineskip 18pt

\begin{center}
{\bf Abstract}
\end{center}
In this paper we have found a classical giant magnon-like solution with both infinite and finite angular 
momentum moving in $Sch_5\times S^5$ with B-field, which is believed to be dual to dipole-deformed
$\mathcal{N}=4$ super Yang-Mills theory. 
This string state propagates as a point particle in non-trivial subspace of the $Sch_5$ space but
shows a giant magnon-like property in the $S^2$ subspace. 
We derive the energy-momentum dispersion
relations and their {\it finite-size correction} for the case of finite but large angular momentum.

\end{titlepage}
%\end{quote}
%\vskip 1cm \centerline{\today}
\newpage
\baselineskip 18pt

%%%%%%%%%%%%%%%%%%%%%%%%%%%%%%%%%%%%%%%%%%%%%%%%%
\def\nn{\nonumber}
%%%%%%%%%%%%%%%%%%%%%%%%%%%%%%%%%%%%%%%%%%%%%%%%%
%%%%%%%%%%%%%%%%%%%%%%%%%%%%%%%%%%%%%%%%%%%%%%%%%%%%%
\def\tr{{\rm tr}\,}
\def\p{\partial}
\newcommand{\non}{\nonumber}
\newcommand{\bea}{\begin{eqnarray}}
\newcommand{\eea}{\end{eqnarray}}
\newcommand{\bde}{{\bf e}}
\renewcommand{\thefootnote}{\fnsymbol{footnote}}
\newcommand{\be}{\begin{eqnarray}}
\newcommand{\ee}{\end{eqnarray}}
%\newcommand{\h}{\hspace{0.5cm}}
%%%%%%%%%%%%%%%%%%%%%%%%%%%%%%%%%%%%%%%%%%%%%%%%%%%%

\vskip 0cm

\renewcommand{\thefootnote}{\arabic{footnote}}
\setcounter{footnote}{0}

%\setcounter{equation}{0}
%%%%%%%%%%%%%5%%%%%%%%%%%%%%%%%%%%%%%%%%%%%%%%%%%%%%
\section{Introduction}
%%%%%%%%%%%%%5%%%%%%%%%%%%%%%%%%%%%%%%%%%%%%%%%%%%%%

Integrability is a key feature in AdS/CFT correspondence to find
exact solutions of both string theory in AdS space and gauge
theory on its boundary. Classical string solutions, such as giant
magnon, are related to asymptotic particle states which describe
certain non-BPS operators of dual conformal field theory (CFT). This special property,
discovered in original formulation of the AdS/CFT, has also
appeared in various deformations of the correspondence.

Dipole deformation of the ${\cal N}=4$ super-Yang-Mills (SYM) theory is one of recent
developments in this direction. With minimal non-locality, imposed
by dipole deformation in a null light-cone direction, the theory
maintains non-relativistic conformal symmetry in three-dimensional
perpendicular directions \cite{Z10,Z11,Z12,Z13}. Furthermore, the dual
supergravity background has been worked out and identified with
geometry known as ``Schr\"odinger'' space-time. It is an
interesting issue if this duality can be established in the
context of integrability, as the original AdS/CFT duality has
shown.

Dipole deformation is  
a special case of integrable Yang-Baxter deformations of the $AdS_5\times S^5$ string \cite{DMV}
as shown in \cite{MY,KKSY}. 
More recently, this issue is addressed again \cite{GLZ} as 
a Drinfeld twist \cite{Z5} of the
${\cal N}=4$ SYM as is the case of other $\star$-product
deformations \cite{Z28}. 
A unified point of view on these integrability structures has been provided in \cite{Stijn}.
Furthermore, in \cite{GLZ} a weak coupling limit in spin chain $sl(2)$ sector has been studied and
semiclassical solutions such as spinning BMN-like strings have been also derived
(see also \cite{ext1}). 

Full quantum integrability of the deformed theory can be encoded in the world-sheet $S$-matrix, which
is conjectured to be Drinfeld-Reshetikhin twist of the undeformed one \cite{GLZ}. 
The giant magnons appear as fundamental asymptotic particles not only in the undeformed theory 
but also in several related theories such as $\beta$-deformed theory \cite{ABBN} and
$\eta$-deformed ${\cal N}=4$ SYM \cite{DMV}. 
Therefore, it is very important to find giant magnon-like solutions in the 
${Sch}_5\times S^5$ string target space along with exact energy-momentum dispersion relation
to understand full non-perturbative AdS/CFT correspondence of the
deformed theory. 
For this purpose, we consider a classical string
configuration which lives both in the Schr\"odinger space and
sphere, whose dispersion relation have a similar form as the
original giant magnon solution of the undeformed theory. In the
supergravity limit, our solution should reproduce the supersymmetric
BMN-like solution.

Another topic we address in this paper is the {\it finite-size} correction to the energy-momentum 
dispersion relation derived from the exact classical string solution with {\it finite angular momentum}.
This information can provide valuable information on the $S$-matrix between the particles, hence 
some insight into non-perturbative aspects of the dipole-deformed theory.

This paper is organized as follows. In section 2 we introduce the metric of
the ``Schr\"odinger'' space-time and impose conformal gauge in the Polyakov string
action along with Virasoro constraints. In section 3 we obtain the giant magnon-like
state and its dispersion relation in the decompactifed limit where the angular momentum in $S^2$ gets infinite. 
Interesting result on the finite-size correction when the angular momentum gets large but finite is
explained in section 4. We conclude the paper in in section 5 with some comments and future research directions.

\setcounter{equation}{0}
%%%%%%%%%%%%%5%%%%%%%%%%%%%%%%%%%%%%%%%%%%%%%%%%%%%%%%%%%%%%%%%%%%%%%%%%%
\section{The string Lagrangian and Virasoro constraints}
%%%%%%%%%%%%%5%%%%%%%%%%%%%%%%%%%%%%%%%%%%%%%%%%%%%%%%%%%%%%%%%%%%%%%%%%%
According to \cite{GLZ}, the metric on $Sch_5\times S^5$ in
global coordinates can be written as \bea\label{gm}
\frac{ds^2}{l^2}=-\left(1+\frac{\mu^2}{Z^4}\right)dt^2+ \frac{2 dt
dV-\vec{X}^2 dt^2+d\vec{X}^2+dZ^2}{Z^2}
+\left(d\psi+P\right)^2+ds^2_{CP^2},\eea where the metric on $S^5$
is given as $U(1)$ Hopf fibre over $CP^2$. 
The parameter $\mu$ is deforming the background from the usual AdS space. 
The $B$-field is given
by \bea\label{B} \alpha' B=l^2 \frac{\mu}{Z^2}
dt\wedge\left(d\psi+P\right).\eea
The metric of $CP^2$ and Hopf fibre over it can be expressed by \cite{ext1}
\bea
ds_{CP^2}^2&=&d\theta_1^2+\frac{1}{4}\sin^2\theta_1\left[
\cos^2\theta_1(d\phi_1+\cos\theta_2 d\phi_2)^2+d\theta_2^2+\sin^2\theta_2 d\phi_2^2
\right],\\
P&=&\frac{1}{2}\sin^2\theta_1 (d\phi_1+\cos\theta_2 d\phi_2).
\eea
We focus on a $S^2$ subspace of $S^5$ by choosing
\bea
\phi_1=0,\qquad \theta_2=\frac{\pi}{2}
\eea
which lead to
\bea
P=0;\qquad ds_{CP^2}^2\quad\to\quad ds^2=d\theta_1^2+\frac{1}{4}\sin^2\theta_1 d\phi_2^2.
\eea
We further consider string solutions in a
subspace of $Sch_5$ obtained by $\vec{X}=0,\ Z=Z_0$.
Also we set $l=1,\alpha'=1$ from now on.
The resulting metric is then simplified to ($\theta\equiv\theta_1, \phi\equiv\phi_2/2$)
\bea
\label{sub} &&ds^2=g_{MN}dX^MdX^N
=-\left(1+\frac{\mu^2}{Z_0^4}\right)dt^2+ \frac{2}{Z_0^2}dt dV
+d\psi^2+d\theta^2+\sin^2\theta d\phi^2,
\\ \nn && B=b_{MN}dX^MdX^N=\frac{\mu}{Z_0^2}
dt\wedge d\psi.
\eea 

In our considerations we will use conformal gauge in the Polyakov
string action, in which the string Lagrangian and Virasoro
constraints have the form 
\bea
\label{l}
&&\mathcal{L}_s=\frac{T}{2}\left(G_{00}-G_{11}+2B_{01}\right) \\
\label{00} && G_{00}+G_{11}=0, \\ &&\label{01} G_{01}=0,
\eea 
where the induced metric and $B$ fields are given by
\bea\nn G_{mn} = g_{MN}\p_m X^M\p_nX^N,\h B_{mn} = b_{MN}\p_m
X^M\p_nX^N,\h M,N = (0,1,\ldots,9),
\eea
with the derivatives w.r.t. the world-sheet coordinates $\eta^0=\tau,\ \eta^1=\sigma$.

We consider the following
string embedding in the background, as given in (\ref{sub}),
\bea\label{se} 
t=\kappa\tau,\h V=\mu^2 m \tau,\h \psi=\omega_1
\tau,\h \phi=\omega\tau+f(\xi),\h \theta=\theta(\xi),\h
\xi=\sigma-v\tau.
\eea 
This choice implies that the string behaves like a point particle in the
subspace of $Sch_5$ while the stringy behaviour appears only in the subspace 
$S^2$. 
In a particular limit where the string image disappears in the $S^2$, the 
configuration is reduced to the spinning BMN-like solution considered in \cite{GLZ}.
This condition fixes the constant coordinate $Z_0$ to 
\bea
Z=Z_0=\sqrt{\frac{\kappa}{m}}.
\eea

Replacing (\ref{se}) into (\ref{l}) one finds that the string
Lagangian is reduced to an effective one-dimensional one (prime is used for $d/d\xi$) 
\bea\label{Lr}
L=-\frac{T}{2}
\left\{(1-v^2)\theta'^2+\kappa^2-\beta^2\omega^2
+\left[(1+v)f'-\omega\right]\left[(1-v)f'+\omega\right]\sin^2\theta\right\},\eea
where the deformation parameter $\beta$ is defined by
\bea\label{kappaeff}
\beta^2\equiv\frac{\omega_1^2+\mu^2m^2}{\omega^2}.
\eea
The equation of motion from (\ref{Lr}) gives a solution for $f$
\bea\label{fif} 
f'(\xi)=\frac{1}{1-v^2}\left(\frac{C}{\sin^2\theta}-v\omega\right),
\eea 
where $C$ is an arbitrary integration constant.

From (\ref{se}) and (\ref{fif}), the
Virasoro constraints (\ref{00}), (\ref{01}) can be written as
\bea\label{00r}
\theta'^2=\frac{1}{1+v^2}\left[\kappa^2-\beta^2\omega^2+\frac{4C v \omega}{(1-v^2)^2}- \frac{(1+v^2)(C^2
\csc^2\theta+\omega^2 \sin^2\theta)}{(1-v^2)^2}\right],\eea
\bea\label{01r} C \omega=v(\kappa^2-\beta^2\omega^2).\eea
The first Virasoro constraint (\ref{00r}) is equivalent to
the first integral of the equation of motion for $\theta$ as shown
in general case \cite{PB2012}.
The second constraint determines the integration constant $C$.
After replacing it in (\ref{00r}) one finds that the non-isometric coordinate $\theta$ satisfies 
a first-order ordinary differential equation:
\bea\label{00rr} \theta'^2 = \frac{\left(\kappa^2-\beta^2\omega^2-\omega^2
\sin^2\theta\right)\left(\omega^2 \sin^2\theta-
v^2(\kappa^2-\beta^2\omega^2)\right)}{(1-v^2)^2 \omega^2
\sin^2\theta} .\eea 

\setcounter{equation}{0}
%%%%%%%%%%%%%5%%%%%%%%%%%%%%%%%%%%%%%%%%%%%%%%%%%%%%%%%%
\section{The giant magnon-like solution in infinite volume}
%%%%%%%%%%%%%5%%%%%%%%%%%%%%%%%%%%%%%%%%%%%%%%%%%%%%%%%%
\subsection{Solution}
The giant magnon is introduced in \cite{HM06} as a string image 
on $S^2$, which is dual to the magnon excitation of the SYM spin chains.
The geometric meaning of the momentum carried by the magnons is a deficit
angle of $\phi$ of the infinite-size string on the equator $\theta=\pi/2$ of the $S^2$ space. 
Therefore, we impose the following condition on a giant magnon-like solution
\bea\label{isc}
\theta'^2 =0\h \mbox{for}\h \theta=\frac{\pi}{2},
\eea
only when we consider the giant magnon in infinite volume.
For the case of finite volume, this condition should be relaxed as we will see in next section.

As can be seen from (\ref{00rr}), this condition reduces to 
\bea 
\left(\kappa^2-\beta^2\omega^2-\omega^2\right)\left(\omega^2-
v^2(\kappa^2-\beta^2\omega^2)\right)=0.
\label{condi}
\eea 
Among two possibilities, we choose for the giant magnon
\bea 
\kappa^2=(1+\beta^2)\omega^2.
\label{deform}
\eea 
This is consistent with the undeformed case $\beta=0$ ($\omega_1=0,\ \mu=0$) which leads to
$\kappa=\omega$.
Along with Eq.(\ref{01r}), this also fixes $C=v\omega$.

With this choice, Eq.(\ref{00rr}) simplifies to 
\bea\label{intm} 
\theta'^2 = \frac{\omega^2 \cos^2\theta\ (\sin^2\theta-v^2)}{(1-v^2)^2
\sin^2\theta},\qquad
f'(\xi)=\frac{v\omega}{1-v^2}\left(\frac{1}{\sin^2\theta}-1\right).
\eea
The solution of this equation is given by
\bea\label{iss} 
\cos\theta(\xi)=\sqrt{1-v^2}\ {\rm sech}\left(\frac{\omega}{\sqrt{1-v^2}}\xi\right).
\eea
This is exactly the Hofman-Maldacena solution for the
infinite-size giant magnon \cite{HM06}. Replacing (\ref{iss}) into
(\ref{fif}), one can find the solution for the isometric
coordinate $\phi$ on $S^2$.

%%%%%%%%%%%%%5%%%%%%%%%%%%%%%%%%%%%%%%%%%%%%%%%%%%%%
\subsection{The energy-charge relation}
%%%%%%%%%%%%%5%%%%%%%%%%%%%%%%%%%%%%%%%%%%%%%%%%%%%%
For the case under consideration the conserved quantities corresponding to isometric 
coordinates $t$, $V$,
$\psi$ and $\phi$ are the
string energy $E_s$, spin $M$ and two angular momenta - $J_1$ and $J$. 
In the limit of decompactified string, with $L\to\infty$, 
these conserved charges are given by
\bea\label{CC} 
&&E_s=\int_{-L/2}^{L/2}d\sigma\frac{\partial{\cal L}_s}{\partial(\partial_0 t)}=
 T\kappa \int_{-L/2}^{L/2}d\sigma=T\kappa L, \\
\nn &&M=\int_{-L/2}^{L/2}d\sigma\frac{\partial{\cal L}_s}{\partial(\partial_0 V)}
= T m \int_{-L/2}^{L/2}d\sigma=T m L, \\ 
\nn &&J_1=\int_{-L/2}^{L/2}d\sigma\frac{\partial{\cal L}_s}{\partial(\partial_0 \psi)}= T \omega_1
\int_{-L/2}^{L/2}d\sigma=T \omega_1 L, \\ \nn 
&&J=\int_{-L/2}^{L/2}d\sigma\frac{\partial{\cal L}_s}{\partial(\partial_0 \phi)}= T \omega\left[
\int_{-L/2}^{L/2}d\sigma- \frac{1}{(1-v^2)}\int_{-L/2}^{L/2}\cos^2\theta d\sigma
\right]= T\left[\omega L-2\sqrt{1-v^2} \right].
\eea

While each of these quantities diverges, a finite combination is possible if we consider
\bea\label{ecr1} 
E_s&-&\sqrt{\mu^2 M^2+J_1^2+J^2}\\ \nn
&=&T\kappa L-\sqrt{(\mu TmL)^2+(T \omega_1 L)^2+(TL\omega)^2
\left[1-\frac{2\sqrt{1-v^2}}{L\omega}\right]^2}\\ \nn
&=&TL\left\{\kappa-\sqrt{1+\beta^2}\ \omega
+\frac{2}{\sqrt{1+\beta^2}}\frac{\sqrt{1-v^2}}{L}\right\}\\ \nn
&=&\frac{2T}{\sqrt{1+\beta^2}}\ \sqrt{1-v^2},\nn
\eea 
where we used Eq.(\ref{deform}) at the last line.

To establish correspondence with the dual dipole-deformed SYM theory, we compute
the angle deficit 
\bea\label{ad} 
\Delta\phi= \int_{-\infty}^{\infty} f'(\xi) d\xi= 
\arccos v\quad
\to\quad
v=\cos\frac{ \Delta\phi}{2}.
\eea 
Identifying $\Delta\phi$ with the momentum $p$ of the magnon excitation,
we can establish the energy-momentum dispersion relation 
as follows:
\bea\label{ecrf}  
E_s-\sqrt{\mu^2 M^2+J_1^2+J^2}
=\frac{2T}{\sqrt{1+\beta^2}}\ \sin\frac{p}{2}.
\eea
This result shows how the deformation affects on the energy-momentum dispersion relation of 
the giant magnon-like string state in $Sch_5\times S^5$.
In the undeformed limit $\beta=0$ ($\omega_1=0,\ \mu=0$), this reduces to 
that of ordinary giant magnon.
In the point-particle limit ($p=0,\ J=0$), this relation reduces to that of
spinning BMN-like strings considered in \cite{GLZ}.

\setcounter{equation}{0}
\section{The giant magnon-like solution in finite volume}
%%%%%%%%%%%%%5%%%%%%%%%%%%%%%%%%%%%%%%%%%%%%%%%%%%%%%%%%
\subsection{Solution}
Introducing new variables 
\bea\label{ratios}
\chi=\cos^2\theta,\qquad W=\frac{\kappa^2}{\omega^2}-\beta^2,
\eea
we can rewrite Eq.(\ref{00rr}) as
\bea\label{ecp} 
\chi'^2 =\frac{4 \omega^2}{(1-v^2)^2} \chi
(\chi_p-\chi)(\chi-\chi_m),\eea where 
$\chi_m$ and $\chi_p$ are given by 
\bea\label{roots}
&&\chi_p=1-v^2W,\qquad \chi_m=1-W.\eea

The solution of Eq.(\ref{ecp}) is then given by
\bea\label{sxi} \xi(\chi)&=&
\frac{1-v^2}{2\omega} \int
\frac{d\chi}{\sqrt{\chi(\chi_p-\chi)(\chi-\chi_m)}} \\ \nn &=&
\frac{1-v^2}{\omega} \frac{1}{\sqrt{\chi_p}}
\mathbf{F}\left(\arcsin\sqrt{\frac{\chi_p-\chi}{\chi_p-\chi_m}},
1-\frac{\chi_m}{\chi_p}\right),\eea where $\mathbf{F}$ is the
incomplete elliptic integral of first kind.
Replacing (\ref{sxi}) into (\ref{fif}), one can find the solution
for the isometric coordinate $\phi$ on $S^2$.

%%%%%%%%%%%%%5%%%%%%%%%%%%%%%%%%%%%%%%%%%%%%%%%%%%%%%%%%%%%%%%%%%%%%%%%%%
\subsection{The conserved quantities}

For the case at hand, the conserved quantities corresponding to
isometric coordinates $t$, $V$, $\psi$ and $\phi$ are the string
energy $E_s$, spin $M$ and two angular momenta - $J_1$ and
$J$ as introduced in (\ref{CC}). 
By changing the integration variable from $d\xi$ to $d\chi/\chi'$
for the finite size $L$, we can express the charges by
\bea\label{Es} E_s&=&
2T\frac{(1-v^2)\kappa}{\omega\sqrt{\chi_p}}\ \mathbf{K}(1-\epsilon),\\
\label{S} M&=& 2T\frac{(1-v^2)m}{\omega\sqrt{\chi_p}}\
\mathbf{K}(1-\epsilon),\\
\label{Jp} J_1&=&
2T\frac{(1-v^2)\omega_1}{\omega\sqrt{\chi_p}}\ \mathbf{K}(1-\epsilon),\\
\label{Jf} J&=&
2T\sqrt{\chi_p}\left[\mathbf{K}(1-\epsilon)-\mathbf{E}(1-\epsilon)\right].\eea
While these charges diverge as $\mathbf{K}\to\infty$ in large volume limit, 
the ratios between them remain finite.

The angular difference $\Delta\phi$ can be also obtained as 
\bea\label{df} \Delta\phi=
\frac{2v}{\sqrt{\chi_p}}\left[\frac{1}{v^2} \
\mathbf{\Pi}\left(-\frac{\chi_p}{1-\chi_p}(1-\epsilon),1-\epsilon\right)-
\mathbf{K}(1-\epsilon)\right].\eea Here $ \mathbf{K}(1-\epsilon)$,
$\mathbf{E}(1-\epsilon)$ and
$\mathbf{\Pi}\left(-\frac{\chi_p}{1-\chi_p}(1-\epsilon),1-\epsilon\right)$
are the complete elliptic integrals of first, second and third
kind, and 
\bea\label{ElMod} \epsilon=\frac{\chi_m}{\chi_p}.\eea

%%%%%%%%%%%%%5%%%%%%%%%%%%%%%%%%%%%%%%%%%%%%%%%%%%%%%%%%%%%%%%%%%%%%%%%%%
\subsection{The energy-charge relation}
%%%%%%%%%%%%%5%%%%%%%%%%%%%%%%%%%%%%%%%%%%%%%%%%%%%%%%%%%%%%%%%%%%%%%%%%%
From the explicit expressions for the charges,
the energy dispersion relation of the giant magnon  
given in (\ref{ecrf}) can be expressed by
\bea\label{ecr1} &&E_s-\sqrt{\mu^2M^2+J_1^2+J^2}=
\\ \nn 
&&
2T\ \frac{(1-v^2)}{\sqrt{1-v^2W}}\ \mathbf{K}(1-\epsilon)
\left[\sqrt{\beta^2+W}-
\sqrt{\beta^2+
\left[\frac{1-v^2W}{1-v^2}\left(1-\frac{\mathbf{E}(1-\epsilon)}{\mathbf{K}(1-\epsilon)}
\right)\right]^2}\right].\eea

Now we consider the energy correction for large but finite $L\gg T$.
Since the ratio $\mathbf{E}/\mathbf{K}$ is very small in this limit, we expand
first 
\bea\label{ecr2} &&E_s-\sqrt{\mu^2M^2+J_1^2+J^2}\approx
2T\frac{(1-v^2)}{\sqrt{1-v^2W}}\times
\\ \nn 
&&
\left\{
\left[\sqrt{\beta^2+W}-
\sqrt{\beta^2+\left(\frac{1-v^2W}{1-v^2}\right)^2}\right]
\mathbf{K}(1-\epsilon)+
\frac{\left(\frac{1-v^2W}{1-v^2}\right)^2}{\sqrt{\beta^2+\left(\frac{1-v^2W}{1-v^2}\right)^2}}\mathbf{E}(1-\epsilon)
\right\}.\eea
Now we assume that the parameters behave as follows for small $\epsilon$:
\bea\label{ep}
&&v=v_0+(v_1+v_2\log\epsilon)\epsilon,
\\ &&W=W_0+(W_1+W_2\log\epsilon)\epsilon.\eea
From conditions $\Delta\phi=p$ in (\ref{df}) and (\ref{ElMod}) with (\ref{roots}), 
one can find the coefficients as
\bea\label{v}
v_0&=&\cos\frac{p}{2},\h
v_1=\frac{1-\log 16}{4}\cos\frac{p}{2}\sin^2\frac{p}{2},\h
v_2=\frac{1}{4}\cos\frac{p}{2}\sin^2\frac{p}{2},\\
W_0&=&1,\h W_1=-\sin^2\frac{p}{2},\h W_2=0.\eea
The coefficient of $\mathbf{K}(1-\epsilon)$ in (\ref{ecr2}) is as small as ${\cal O}(\epsilon)$ so that
the logarithmic divergent term disappears.

With these and expansions of the elliptic functions, we find 
\bea\label{finiteE} E_s-\sqrt{\mu^2M^2+J_1^2+J^2}=
\frac{2T\sin\frac{p}{2}}{\sqrt{1+\beta^2}}\left[
1-\frac{\sin^2\frac{p}{2}+\beta^2\left(1-5\cos^2\frac{p}{2}\right)}{4(1+\beta^2)}\ \epsilon+{\cal O}(\epsilon^2)\right].
\eea
The series expansion of $J$ for small $\epsilon$ is
\bea
J\approx T\sin\frac{p}{2}\left(-2-\log\frac{\epsilon}{16}\right),
\eea
from which the expression for $\epsilon$ for $J\gg T$ can be found as
\bea\label{e} \epsilon= 16
\exp\left(-\frac{J}{T\sin\frac{p}{2}}-2\right).\eea
Combining together, we obtain the leading finite-size correction of the energy-charge
dispersion relation to be
\bea\label{finiteEf} E_s-\sqrt{\mu^2M^2+J_1^2+J^2}=
\frac{2T\sin\frac{p}{2}}{\sqrt{1+\beta^2}}\left[
1-4\ \frac{\sin^2\frac{p}{2}+\beta^2\left(1-5\cos^2\frac{p}{2}\right)}{1+\beta^2}
\ e^{-\frac{J}{T\sin\frac{p}{2}}-2}\right],
\eea
where the deformation parameter $\beta$ is defined in Eq.(\ref{ecrf}).
The leading term is the energy dispersion relation in infinite volume as we have obtained in the previous
section. The second term is the finite-size correction to the energy and is our main result in this paper.
The coefficient in front of the exponential factor and its dependence on the deformation parameter $\beta$
defined in (\ref{kappaeff}) contains important information on the interaction between the giant magnon states.

For the undeformed case of $\beta\to 0$, this result reduces to
\bea\label{finiteEfnodeform} E_s-J=
2T\sin\frac{p}{2}\left[
1-4\sin^2\frac{p}{2}\ e^{-\frac{J}{T\sin\frac{p}{2}}-2}\right],
\eea
which was obtained previous in \cite{FiniteGM}.

\setcounter{equation}{0}
%%%%%%%%%%%%%5%%%%%%%%%%%%%%%%%%%%%%%%%%%%%%%%%%%%%%
\section{Concluding remarks}
%%%%%%%%%%%%%5%%%%%%%%%%%%%%%%%%%%%%%%%%%%%%%%%%%%%%

In this paper, we have computed classical giant magnon-like solutions moving in the $Sch_5\times S^5$ target space.
We have considered the string configuration similar to conventional giant magnons, namely, 
point-like in the $Sch_5$ space and extended string-like in the $S^2$ 
subspace of $S^5$.
We have obtained results for both infinite angular momentum and large but finite one.
The conserved charges and the corresponding energy-charge relations are expressed in terms of elliptic
integrals. 
We have confirmed that these results are consistent with previously known results in point-like and 
undeformed limits.
A possible generalization is to consider dyonic giant magnon solution in $Sch_5\times S^5$.
This solution can live in $S^3$ where additional finite angular momentum is introduced.
Another direction is to deform the sphere $S^5$ in addition to the dipole deformation of the $Sch_5$.
It will be important to see if two or more deformations can maintain (classical) integrability.
We hope to report on these in near future.

{\bf Note added:} After the first version of this paper was posted in the {\tt arXiv}, 
new paper has appeared where classical string solutions have derived \cite{Zoakos}. 
Main difference from ours is that it considers string-like solutions even in the $Sch_5$ space.

\section*{Acknowledgements}
This work is supported in part by the Brain Pool program 171S-1-3-1765
from KOFST and by National Research Foundation of Korea (NRF) grant
(NRF-2016R1D1A1B02007258) (CA) and
the NSF grant DFNI T02/6 (PB). The authors would like to thank E. Colgain and R. C. 
Rashkov for the useful discussions. CA also thanks ICTP for a support through Senior Associates program.

\end{document}